\documentclass[reprint,showkeys,citeautoscript]{revtex4-2}
\setcitestyle{super}

\usepackage{amsmath}
\usepackage{amssymb,bm}
\usepackage{graphicx,color}

\newcommand{\e}{\mathrm{e}}
\newcommand{\ii}{\mathrm{i}}
\newcommand{\dd}{\,\mathrm{d}}

\newcommand{\eq}{eq}
\newcommand{\eqs}{eqs}

\makeatletter
\renewcommand{\fnum@figure}{\textbf{Figure~\thefigure}}
\makeatother

\renewcommand{\thetable}{\arabic{table}}
\makeatletter
\renewcommand{\fnum@table}{\textbf{Table~\thetable}}
\makeatother

\renewcommand{\thesection}{\arabic{section}}
\renewcommand{\thesubsection}{\arabic{section}.\arabic{subsection}}

\begin{document}
\title{Photon Correlations Control Quantum Fisher Information in Coherent Multidimensional Spectroscopy}

\author{Chenghong Huang}
\affiliation{Department of Chemistry, The University of Tokyo, Tokyo 113-0033, Japan}
\author{Yuta Fujihashi}
\affiliation{Department of Chemistry, The University of Tokyo, Tokyo 113-0033, Japan}
\author{Akihito Ishizaki}
\email{ishizaki@chem.s.u-tokyo.ac.jp}
\affiliation{Department of Chemistry, The University of Tokyo, Tokyo 113-0033, Japan}

\begin{abstract}
Coherent multidimensional spectroscopy probes interactions and dynamics of molecular systems through nonlinear optical signals. Quantum light excitation has been explored as a means of controlling these signals, raising the question of how excitation-light correlations affect the precision of molecular parameter estimation from the emitted signal. In this work, we show that photon-number correlations and two-photon coherence play distinct roles in encoding molecular information in the emitted rephasing signal. In the weak-emission regime and within the single Schmidt pair approximation, we derive a normally ordered information law in which the joint photon-number moment and the squared magnitude of the two-photon coherence weight the amplitude and phase sensitivities of the molecular response, respectively. For matched temporal modes and mean photon numbers in each input mode, two-mode squeezed vacuum enhances both weights relative to coherent excitation, whereas independent displaced squeezing preserves the amplitude weight but reduces the phase weight. The background population and its parameter dependence also affect the total quantum Fisher information, which therefore cannot be inferred from the coherent mean-field signal alone. We derive a leading-order bound on its enhancement over coherent excitation when the background population is nonnegative and common to the compared sources. These findings provide criteria for selecting excitation sources according to how the target molecular parameter affects the amplitude and phase of the response.
\end{abstract}

\keywords{\it quantum-light spectroscopy, quantum Fisher information, two-dimensional electronic spectroscopy, photon correlations, quantum metrology}

\maketitle

\section{Introduction}
Coherent multidimensional spectroscopy correlates excitation and detection frequencies across a controllable waiting time~\cite{hybl1998, jonas2003, brixner2004, cowan2004}. Specifically, two-dimensional (2D) electronic spectroscopy has resolved electronic couplings and energy transfer dynamics in photosynthetic complexes~\cite{brixner2005, zigmantas2006, ginsberg2009, schlau2011} and characterized oscillatory dynamics associated with electronic and vibronic coherences~\cite{engel2007, collini2010, romero2014, thyrhaug2018}. Theoretical descriptions of nonlinear optical responses and exciton dynamics relate these spectra to molecular parameters such as electronic couplings, reorganization energies, and transfer rates~\cite{mukamel1995, abramavicius2009, ishizaki2009}. In quantitative applications, these parameters are inferred by comparing measured spectra with theoretical simulations~\cite{brixner2005,read2008,hayes2011} or through global and target analysis~\cite{van2004,volpato2016}. Such inference constitutes a parameter estimation problem, motivating the assessment of how precisely molecular parameters can be estimated from optical signals, beyond the strength of spectral features alone~\cite{bolzonello2024}.

Quantum light introduces photon statistics and intermode correlations as additional controls over nonlinear optical responses~\cite{dorfman2016, schlawin2018}. Quantum-correlated light has enabled sub-shot-noise absorption spectroscopy~\cite{tapster1991su, matsuzaki2022su, matsuzaki2024su}, infrared spectroscopy with visible-light detection~\cite{kalashnikov2016, mukai2021qu, lindner2021no}, and fluorescence lifetime measurements without pulsed lasers~\cite{fujihashi2020ge, li2023si, harper2023en, alvarezMendoza2025, li2025co}. At low photon flux, entangled two-photon absorption exhibits a correlation-induced contribution that scales linearly with photon flux~\cite{javanainen1990, lee2006en}. Temporal, spectral, and photon-number correlations have also been explored for time-resolved nonlinear spectroscopy~\cite{roslyak2009no, raymer2013en, dorfman2014st, ishizaki2020probing, chen2022mo, fujihashi2023, fujihashi2026} and pathway selectivity~\cite{fujihashi2024, kizmann2023, yadalam2023, yadalam2024, kim2024pr, hong2025}. To overcome the weak nonlinear signals associated with isolated photon pairs~\cite{landes2021, raymer2021, landes2021exp}, bright correlated beams have been investigated in nonlinear spectroscopy~\cite{eto2021en, raymer2022th, fujihashi2023, hashimoto2024, eto2025sp, fan2026qu, jadoun2025, jadoun2026, eto2026al}. These developments demonstrate how nonclassical correlations can modify spectroscopic signals and motivate examining their implications for molecular-parameter estimation.

Quantum metrology provides a natural framework for this question \cite{boixo2008qu,barts2026qu}. The quantum Fisher information (QFI) of an output state bounds the precision of any measurement through the quantum Cram\'er-Rao bound and depends only on the parameter dependence of that state~\cite{braunstein1994, paris2009, giovannetti2011}. Quantum electrodynamic formulations of nonlinear spectroscopy describe the emitted radiation as the output of light--matter quantum maps acting on broadband temporal modes~\cite{schlawin2025}, allowing different quantum excitation states to be compared within a common mode basis. Quantum metrological analyses have established precision limits for spectroscopy with pulsed quantum light~\cite{munoz2021, albarelli2023, khan2024} and have recently been extended to stimulated Raman and pump-probe spectroscopy~\cite{schlawin2026srs, schlawin2026pp}. In pump-probe spectroscopy, for example, squeezing and pump-probe correlations can improve estimation when the nonclassical optical modes themselves remain part of the detected output~\cite{schlawin2026pp}. In 2D electronic spectroscopy, the quantized excitation fields and the phase-matched emitted signal occupy distinct optical modes. How excitation field correlations determine the molecular information encoded in the emitted signal state remains to be clarified. This raises a practical question for quantum-light spectroscopy: which properties of the excitation field should guide source selection for estimating a given molecular parameter? A parameter can affect both the amplitude and phase of the molecular response, and these two sensitivities need not benefit from the same optical correlations. 

Here we investigate how interaction-selected two-photon correlations determine the QFI of the emitted rephasing signal in 2D electronic spectroscopy. In the weak-emission regime, two selected field moments govern the signal coherence and the joint two-input contribution to the signal population. We derive a normally ordered information law that relates amplitude and phase sensitivities to joint photon-number correlations and two-photon coherence, respectively. 
We also establish a correlation law for the eigenvector QFI enhancement factor, which is independent of the molecular parameter being estimated, together with a leading-order bound on the total QFI enhancement in the presence of a common, nonnegative background population. Within the single Schmidt pair approximation, matched sources with nonzero pair coherence yield identical normalized mean-field 2D electronic spectra, yet encode different amounts of QFI in the emitted quantum state. Numerical calculations for a coupled exciton dimer under coherent, two-mode squeezed vacuum, and independent displaced squeezed excitation confirm these predictions across eight molecular parameters and reveal their consequences for information redistribution and low-brightness scaling.

\section{METHODS}
\label{sec:framework}

\begin{figure*}[t]
    \centering
    \includegraphics{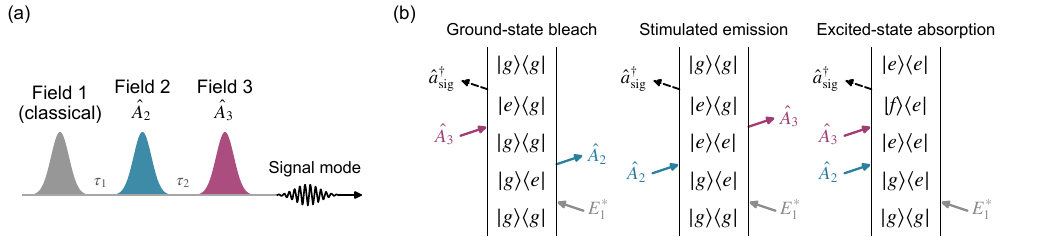}
    \caption{
    Optical configuration and rephasing pathways in the quantum-map formulation.
    (a) Rephasing pulse sequence with a classical first excitation field (gray), quantized second and third fields with mode operators $\hat{A}_2$ (blue) and $\hat{A}_3$ (magenta), and the emitted signal mode. The delays $\tau_1$ and $\tau_2$ separate the three excitation pulses.
    (b) Representative double-sided Feynman diagrams for the ground-state bleach (GSB), stimulated emission (SE), and excited-state absorption (ESA) pathways. Time runs upward; the left and right vertical lines represent the ket and bra sides of the density operator. The labels $g$, $e$, and $f$ denote the ground, one-exciton, and two-exciton states, respectively. Solid arrows indicate interactions with the classical field amplitude $E_1^{\ast}$ (gray) and with the quantized modes $\hat{A}_2$ (blue) and $\hat{A}_3$ (magenta); dashed arrows indicate emission into the signal mode through $\hat{a}_{\rm sig}^{\dagger}$. The operators acting on the ket and bra sides define the pair $(\hat{L}_\ell, \hat{R}_\ell)$ of each pathway.
    }
    \label{fig:scheme}
\end{figure*}

\subsection{Effective emitted signal and selected field moments}
\label{sec:signal-moments}
We consider the estimation of a molecular parameter $\theta$, such as the electronic coupling $J$, the transfer rate $k_{12}$, or the bath relaxation rate $\gamma$. We focus on the rephasing signal in the phase-matching direction $\mathbf{k}_{\rm R} = - \mathbf{k}_1 + \mathbf{k}_2 + \mathbf{k}_3$ in the BOXCARS geometry, which corresponds to the photon-echo signal~\cite{brixner2004}. We treat field~1 as a prescribed classical drive, and fields~2 and~3 are quantized to retain the two-mode correlations selected by the rephasing interaction. Each quantized field is represented by a single broadband temporal mode with annihilation operator $\hat{A}_j$ ($j = 2, 3$), given by the dominant Schmidt pair of the two fields (Section~\ref{app:modes} of the Supporting Information). The emitted field is projected onto a single retained signal mode with annihilation operator $\hat{a}_{\rm s}$ and Fock states $\lvert n \rangle$.

The excitation source $s$ is chosen from coherent light ($s = \text{coh}$), two-mode squeezed vacuum ($s= \text{TMSV}$), and independently displaced squeezed light ($s=\text{sq}$). The excitation fields are prepared in the state $\hat{\rho}_{\rm exc}^{(s)}$, and signal mode is initially in the vacuum $\lvert 0 \rangle$. The emitted field is obtained by applying the third-order quantum map and tracing out the excitation fields,
\begin{equation}
    \hat{\rho}_{\rm sig}^{(s)}(\theta)
    =
    \operatorname{Tr}_{\rm exc}
    \left[
        \e^{\mathcal{K}_3(\theta)}
        \left(
            \hat{\rho}_{\rm exc}^{(s)}
            \otimes
            \lvert 0 \rangle\langle 0 \rvert
        \right)
    \right].
    \label{eq:rho-sig}
\end{equation}
The third-order generator $\mathcal{K}_3 (\theta)$ is constructed from six double-sided Feynman diagrams, comprising the ground-state bleach (GSB), stimulated emission (SE), and excited-state absorption (ESA) pathways and their Hermitian-conjugate counterparts (Figure~\ref{fig:scheme} and \eq~\ref{eq:K3_appendix} of the Supporting Information). We label $\ell \in \{\mathrm{GSB}, \mathrm{SE}, \mathrm{ESA}\}$ the three pathways that create a signal photon (Figure~\ref{fig:scheme}b), and their Hermitian-conjugate counterparts carry $G_{1\ell}^{\ast}(\theta)$ and generate $\rho_{01} = \rho_{10}^{\ast}$. Each displayed pathway $\ell$ is associated with a complex coefficient $G_{1\ell}(\theta)$ and a pair of excitation operators $(\hat{L}_\ell, \hat{R}_\ell)$ acting on the ket and bra sides of the density operator. The coefficients $G_{1\ell}(\theta)$ contain the molecular response, optical-mode overlaps, signal mode projection, light--matter coupling, and the fixed contribution of field~1. A common rescaling of these coefficients sets the interaction scale $\varepsilon$ (\eq~\ref{eq:interaction-scale} of the Supporting Information).

To leading order in $\varepsilon$, the populations in the signal Fock states $\lvert n \ge 2\rangle$ are of order $\varepsilon^{4}$. In this weak-emission regime, we analyze the normalized vacuum--one-photon block of $\hat{\rho}_{\rm sig}^{(s)}(\theta)$, 
\begin{equation}
    \hat{\rho}^{(s)} (\theta)
    = 
    \begin{pmatrix}
        1 - p^{(s)}(\theta)     & \rho_{10}^{(s)\ast} (\theta) 
        \\
        \rho_{10}^{(s)}(\theta) & p^{(s)}(\theta)
    \end{pmatrix},
    \label{eq:signal}
\end{equation}
in the basis $\{\lvert 0 \rangle, \lvert 1 \rangle \}$. The matrix elements $\rho_{10}^{(s)}(\theta) = \langle 1 \rvert \hat{\rho}^{(s)}(\theta) \lvert 0 \rangle$ and $p^{(s)} = \langle 1 \rvert \hat{\rho}^{(s)}(\theta) \lvert 1 \rangle$ denote the vacuum--one-photon coherence and one-photon population, respectively. 

The coherence arises from the first-order term of $\e^{\mathcal{K}_3(\theta)}$ in \eq~\ref{eq:rho-sig}, which gives $\rho_{10}^{(s)}(\theta) = - \ii \sum_\ell G_{1\ell}(\theta) \langle \hat{R}_\ell \hat{L}_\ell \rangle_s$. For all three pathway classes, the product of the two excitation operators is the same, $\hat{R}_\ell \hat{L}_\ell = \hat{A}_2 \hat{A}_3 = \hat{X}$. The rephasing interaction therefore selects $\hat{X}$, and we define the corresponding source moments as 
\begin{align}
    C^{(s)}
    & =
    \langle \hat{X} \rangle_s,
    \label{eq:moments-RP}
    \\
    M^{(s)}
    & =
    \langle\hat{X}^\dagger \hat{X} \rangle_s
    =
    \langle \hat{n}_2 \hat{n}_3 \rangle_s,
    \label{eq:population-moment}
\end{align}
with $\hat{n}_j = \hat{A}_j^\dagger \hat{A}_j$. The coherence then factorizes as
\begin{equation}
    \rho_{10}^{(s)}(\theta)
    =
    \mathcal{G}(\theta)C^{(s)},
    \label{eq:factorization}
\end{equation}
where the molecular response consists of three pathway contributions
\begin{equation}
    \mathcal{G}(\theta)
    =
    \sum_\ell \mathcal{G}_\ell(\theta),
    \qquad
    \mathcal{G}_\ell(\theta) 
    =
    -\ii\, G_{1\ell}(\theta).
    \label{eq:pathway-sum}
\end{equation}

The excitation states and optical modes are independent of $\theta$, so that
\begin{equation}
    \frac{\partial C^{(s)}}{\partial\theta}
    =
    \frac{\partial M^{(s)}}{\partial\theta}
    = 0 .
    \label{eq:theta-independent}
\end{equation}
All dependence of the signal state on $\theta$ therefore enters through the molecular coefficients $G_{1\ell}(\theta)$.

The one-photon population arises from the second-order term of $\e^{\mathcal{K}_3(\theta)}$. Bringing the excitation operators into normal order gives (\eq~\ref{eq:app-population} of the Supporting Information)
\begin{equation}
    p^{(s)}(\theta)
    =
    P_{\rm N}^{(s)}(\theta) + q^{(s)}(\theta)
    \label{eq:population-split}
\end{equation}
with
\begin{align}
    P_{\rm N}^{(s)}(\theta)
    & =
    \lvert \mathcal{G}(\theta) \rvert^{2} M^{(s)}
    \label{eq:population-factorization}
    \\
    q^{(s)}(\theta)
    & =
    b_2(\theta) \langle \hat{n}_2 \rangle_s
    + b_3(\theta) \langle \hat{n}_3 \rangle_s
    + p_\varnothing(\theta)
    \label{eq:q-definition}
\end{align}
where $b_2(\theta)$ and $b_3(\theta)$ are given by $\operatorname{Re}[\mathcal{G}_{\ell}^{\ast}(\theta)\mathcal{G}(\theta)]$ with $\ell = \mathrm{SE}$ and $\mathrm{GSB}$, respectively, and $\mathcal{G}_\ell(\theta)$ is defined in eq~\ref{eq:pathway-sum}. The normally ordered population $P_{\rm N}^{(s)}$ depends on both quantum inputs, whereas the remaining population $q^{(s)}(\theta)$ contains single-mode occupation and input-independent contributions, with $p_\varnothing(\theta) = 0$ in the present quantum map.

The normally ordered contribution can be isolated by comparing four excitation configurations: both fields present ($p_{23}^{(s)}$), only one field present ($p_{2}^{(s)}$, $p_{3}^{(s)}$), and both fields in vacuum ($p_\varnothing$). We define their double difference as 
\begin{equation}
    P_{\rm N}^{(s)} (\theta)
    =
    p_{23}^{(s)}(\theta) - p_{2}^{(s)}(\theta) 
    - p_{3}^{(s)}(\theta) + p_\varnothing(\theta)
    + O(\varepsilon^4).
    \label{eq:population-double-difference}
\end{equation}
Here, removing an input field means replacing it with vacuum while preserving the marginal state of the remaining field and the optical interaction configuration. The subtraction removes contributions depending on either input field separately, while adding back $p_\varnothing$ corrects for the double subtraction of the input-independent contribution. The resulting expression isolates the joint two-input contribution to the signal population at leading order in $\varepsilon$.

For the matched excitation sources considered below, the mean occupations satisfy $\langle \hat{n}_2 \rangle_s = \langle \hat{n}_3 \rangle_s = n$. Equation~\ref{eq:q-definition} then gives
\begin{equation}
    q^{(s)}(\theta)
    =
    n[b_2(\theta)+b_3(\theta)]
    +p_\varnothing(\theta)
    = q(\theta),
    \label{eq:q-matched}
\end{equation}
which is independent of the excitation source to leading order in $\varepsilon$.

\subsection{Quantum Fisher information and information law}
\label{sec:weak-field}
To quantify how sensitively the emitted signal state responds to a small change in a molecular parameter $\theta$, we evaluate its quantum Fisher information (QFI). We suppress the source superscript $s$ until the end of \eq~\ref{eq:weak-QFI-val}. The QFI is defined as~\cite{braunstein1994, paris2009}
\begin{equation}
    F_{\rm Q}(\theta)
    =
    \operatorname{Tr}[\hat{\rho}_{\rm sig}(\theta) \hat{L}_\theta^2],
    \label{eq:qfi}
\end{equation}
where the symmetric logarithmic derivative $\hat{L}_\theta$ is a Hermitian operator satisfying
\begin{equation}
    \frac{\partial }{\partial \theta} 
    \hat{\rho}_{\rm sig}(\theta)
    =
    \frac{1}{2}[\hat{L}_\theta \hat{\rho}_{\rm sig}(\theta) 
    +
    \hat{\rho}_{\rm sig}(\theta) \hat{L}_\theta].
    \label{eq:sld}
\end{equation}
To separate the QFI into contributions from the eigenvalues and eigenvectors of the emitted signal state, we write its spectral decomposition as $\hat{\rho}_{\rm sig}(\theta) = \sum_k \lambda_k(\theta) \lvert k(\theta) \rangle\langle k(\theta) \rvert$. In this eigenbasis, \eq~\ref{eq:sld} leads to
\begin{equation}
    \langle k(\theta) \rvert \hat{L}_\theta \lvert l(\theta) \rangle
    =
    \frac{2}{\lambda_k(\theta) + \lambda_l(\theta)}
    \left\langle k(\theta) \left\vert 
    \frac{\partial }{\partial \theta} \hat{\rho}_{\rm sig}(\theta)
    \right\vert l(\theta) \right\rangle
    \label{eq:sld-elements}
\end{equation}
for $\lambda_k(\theta) + \lambda_l(\theta) > 0$. The diagonal elements are determined by $\partial\lambda_k(\theta) / \partial\theta$, whereas the off-diagonal elements describe changes in the eigenvectors. Substituting \eq~\ref{eq:sld-elements} into \eq~\ref{eq:qfi} gives $F_{\rm Q} = F_{\rm vec} + F_{\rm val}$, where
\begin{equation}
    F_{\rm vec} (\theta)
    =
    2 \sum_{l \neq k}
    \frac{1}{\lambda_l(\theta) + \lambda_k(\theta)}
    \left\lvert 
        \left\langle l(\theta) \left\rvert 
            \frac{\partial}{\partial\theta} \hat{\rho}_{\rm sig}(\theta)
        \right\lvert k(\theta) \right\rangle 
    \right\rvert^2,
    \label{eq:QFI-vec}
\end{equation}
and
\begin{equation}
    F_{\rm val} (\theta)
    =
    \sum_k \frac{1}{\lambda_k(\theta)}
    \left[
        \frac{\partial }{\partial \theta} \lambda_k(\theta)
    \right]^2.
    \label{eq:QFI-val}
\end{equation}
For the weak-emission state in \eq~\ref{eq:signal}, the vacuum--one-photon coherence $\rho_{10}$ is of order $\varepsilon$ and the population $p$ is of order $\varepsilon^{2}$. Evaluating eqs~\ref{eq:QFI-vec} and~\ref{eq:QFI-val} to leading order in $\varepsilon$ (Section~\ref{app:qfi-bound} of the Supporting Information) gives
\begin{gather}
    F_{\rm vec} (\theta)
    =
    4 \left\lvert \frac{\partial }{\partial \theta}
    \rho_{10}(\theta)\right\rvert^{2}
    + O(\varepsilon^4),
    \label{eq:weak-QFI-vec}
\\
    F_{\rm val} (\theta)
    =
    \frac{
        \left[
            \dfrac{\partial}{\partial\theta} (p(\theta) - \lvert \rho_{10}(\theta) \rvert^{2}) 
        \right]^{2}
    }{
        p(\theta) - \lvert \rho_{10}(\theta) \rvert^{2}
    }
    + O(\varepsilon^4).
    \label{eq:weak-QFI-val}
\end{gather}
All ratios below are evaluated to leading order in $\varepsilon$ and are written as equalities. Since the excitation state is independent of the molecular parameter $\theta$, substituting eq~\ref{eq:factorization} into eq~\ref{eq:weak-QFI-vec} gives $F_{\rm vec}^{(s)}(\theta) = 4\lvert C^{(s)}\rvert^{2}\lvert \partial\mathcal{G}(\theta)/\partial\theta\rvert^{2}$. The molecular factor is common to matched sources and cancels in the ratio,
\begin{equation}
    R_{\rm vec}^{(s)}
    =
    \frac{F_{\rm vec}^{(s)}}
         {F_{\rm vec}^{(\rm coh)}}
    =
    \frac{\lvert C^{(s)}\rvert^{2}}
         {\lvert C^{(\rm coh)}\rvert^{2}},
    \label{eq:correlation-law}
\end{equation}
where the coherent reference has matched temporal modes and incident photon-number budget. Thus, the leading enhancement of the eigenvector QFI contribution is determined solely by the interaction-selected excitation correlation.

\paragraph{Normally ordered information law.}
To isolate the contribution of the excitation correlations, we define the normally ordered information $F_{\rm N}^{(s)} (\theta)$ as the QFI of an auxiliary state obtained from \eq~\ref{eq:signal} by replacing $p^{(s)}(\theta)$ with $P_{\rm N}^{(s)}(\theta)$, while retaining the same coherence $\rho_{10}^{(s)}(\theta)$. The smaller eigenvalue of this state is determined to leading order by
\begin{align}
    D^{(s)}(\theta)
    & =
    P_{\rm N}^{(s)}(\theta)
    -
    \lvert \rho_{10}^{(s)}(\theta)\rvert^2
    \nonumber
    \\
    & =
    \lvert \mathcal{G}(\theta) \rvert^2
    (M^{(s)} - \lvert C^{(s)} \rvert^2 ).
    \label{eq:D-definition}
\end{align}
Applying the weak-emission QFI expressions gives
\begin{equation}
    F_{\rm N}^{(s)}(\theta)
    =
    4\lvert C^{(s)}\rvert^2
    \left\lvert
    \frac{\partial}
         {\partial\theta} \mathcal{G}(\theta)
    \right\rvert^2
    +
    \frac{
        \left[
        \dfrac{\partial}
             {\partial\theta}D^{(s)}(\theta)
        \right]^2
    }{
        D^{(s)}(\theta)
    }
    +O(\varepsilon^4).
    \label{eq:FN-two-terms}
\end{equation}
For $D^{(s)}(\theta)>0$, its parameter derivative
satisfies
\begin{equation*}
    \frac{\partial}{\partial\theta} D^{(s)}(\theta)
    =
    2( M^{(s)}-|C^{(s)}|^2 )
    \operatorname{Re}
    \left[
        \mathcal{G}^*(\theta)
        \frac{\partial\mathcal{G}(\theta)}
             {\partial\theta}
    \right].
\end{equation*}
Defining 
\begin{gather}
    \kappa^{(s)} 
    = 
    \frac{\lvert C^{(s)} \rvert^{2}}{M^{(s)}}, \quad
    \chi(\theta) 
    = 
    \arg\left[
        \frac{\partial }{\partial \theta}\ln \mathcal{G}(\theta)
    \right],
\end{gather}
and using the parameter independence of the excitation moments, eq~\ref{eq:FN-two-terms} yields the normally ordered information law
\begin{equation}
    F_{\rm N}^{(s)} (\theta)
    =
    4M^{(s)} 
    \left\lvert 
        \frac{\partial}{\partial\theta}\mathcal{G}(\theta) 
    \right\rvert^{2}
    [
        \cos^2\chi(\theta)
        +
        \kappa^{(s)}\sin^2\chi(\theta)
    ].
    \label{eq:FN-law}
\end{equation}
Here $\kappa^{(s)}$ characterizes the normalized selected correlation, while $\cos^{2} \chi(\theta)$ and $\sin^{2} \chi(\theta)$ represent the amplitude- and phase-encoded parameter sensitivity. 

For a matched coherent reference, $\kappa^{(\rm coh)} = 1$. We define the normally ordered population ratio
\begin{equation}
    Y_{\rm N}^{(s)} 
    =
    \frac{P_{\rm N}^{(s)}(\theta)}{P_{\rm N}^{(\rm coh)}(\theta)}
    =
    \frac{M^{(s)}}{M^{(\rm coh)}},
\end{equation}
which quantifies the relative contribution of the joint two-input excitation to the signal population. This quantity is distinct from the total signal population ratio $p^{(s)}(\theta)/p^{(\rm coh)}(\theta)$, which may contain additional contributions. The corresponding information law is
\begin{equation}
    R_{\rm N}^{(s)}(\theta)
    =
    \frac{F_{\rm N}^{(s)}(\theta)}{F_{\rm N}^{(\rm coh)}(\theta)}
    =
    Y_{\rm N}^{(s)}
    [
        \cos^2\chi(\theta) + \kappa^{(s)} \sin^{2} \chi(\theta)
    ].
    \label{eq:RN-law}
\end{equation}

\paragraph{Total QFI and its bound.}
We now return to the complete signal population $p^{(s)}(\theta) = P_{\rm N}^{(s)}(\theta) + q(\theta)$. Using \eqs~\ref{eq:weak-QFI-val} and~\ref{eq:D-definition}, the smaller eigenvalue of the signal state is $\lambda_-^{(s)}(\theta) = D^{(s)}(\theta) + q(\theta) + O(\varepsilon^4)$. For $q(\theta) > 0$, the total QFI admits the decomposition
\begin{equation}
    F_{\rm Q}^{(s)}(\theta)
    =
    F_{\rm N}^{(s)}(\theta)
    +
    F_{\rm bg}(\theta)
    -
    \Delta_{\rm mix}^{(s)}(\theta)
    +
    O(\varepsilon^4),
    \label{eq:FQ-decomposition}
\end{equation}
where $F_{\rm bg}(\theta) = [\partial q(\theta)/ \partial \theta]^{2} / q(\theta)$. For $D^{(s)}(\theta) > 0$, the mixing correction is
\begin{equation}
    \Delta_{\rm mix}^{(s)}(\theta)
    =
    \frac{D^{(s)}(\theta)q(\theta)}
    {D^{(s)}(\theta)+q(\theta)}
    \left[
        \frac{\partial}{\partial\theta}
        \ln\frac{D^{(s)}(\theta)}{q(\theta)}
    \right]^2
    \ge 0.
    \label{eq:mixing}
\end{equation}
This correction arises from the difference in the relative parameter sensitivities of $D^{(s)} (\theta)$ and $q(\theta)$. For the coherent reference, $D^{(\rm coh)}(\theta) = 0$, so that $\Delta_{\rm mix}^{(\rm coh)}(\theta) = 0$ and eq~\ref{eq:FQ-decomposition} reduces to $F_{\rm Q}^{(\rm coh)}(\theta) = F_{\rm N}^{(\rm coh)}(\theta) + F_{\rm bg}(\theta)$ (eq~\ref{eq:app-FQ-coh} of the Supporting Information). Since the background population $F_{\rm bg}(\theta)$ is common to the matched excitation sources, dividing eq~\ref{eq:FQ-decomposition} by this expression and using $\Delta_{\rm mix}^{(s)} (\theta)\ge 0$ gives the leading-order bound
\begin{equation}
    R_{\rm Q}^{(s)} (\theta)
    =
    \frac{F_{\rm Q}^{(s)} (\theta)}
         {F_{\rm Q}^{(\rm coh)} (\theta)}
    \le
    \frac{
        R_{\rm N}^{(s)}(\theta) + \xi_{\rm bg} (\theta)
    }{
        1+\xi_{\rm bg} (\theta)
    },
    \label{eq:RQ-bound}
\end{equation}
where $\xi_{\rm bg} (\theta) = F_{\rm bg}(\theta) / F_{\rm N}^{(\rm coh)}(\theta) \ge 0$. The upper bound does not exceed $R_{\rm N}^{(s)}(\theta)$ when $R_{\rm N}^{(s)}(\theta) \ge 1$, and does not exceed unity when $R_{\rm N}^{(s)}(\theta) \le 1$.

The interaction-selected information law therefore provides an upper bound on the total QFI enhancement in the presence of common, nonnegative population contributions. This bound holds to leading order for positive collective signal states and does not require the absolute magnitude or microscopic origin of the additional population to be specified. The detailed derivation and validity conditions are given in Section~\ref{app:qfi-bound} of the Supporting Information.

\subsection{Source-state moments}
\label{sec:source-moments}
We compare coherent, two-mode squeezed vacuum (TMSV), and independent displaced squeezed excitation at a fixed mean photon number $n$ per input mode, i.e., at a total mean photon number $\bar{n} = 2n$, so that their differences arise solely from the photon statistics. 

For coherent excitation, each input mode is prepared in a coherent state with mean occupation $n$. For TMSV, $n = \sinh^{2} r$, where $r$ is the two-mode squeezing parameter. For independent displaced squeezed states, the mean occupation is $n = n_{\rm disp} + n_{\rm sq}$, with $n_{\rm disp} = \lvert \alpha_j \rvert^{2}$ and $n_{\rm sq}=\sinh^{2} r$ denoting the displacement and squeezing contributions, respectively. The interaction-selected moments are summarized in Table~\ref{tab:source-moments}.

\begin{table*}
    \centering
    \caption{\bf Interaction-selected Moments for the Matched Excitation Sources.}
    \label{tab:source-moments}
    \renewcommand{\arraystretch}{1.5}
    \begin{tabular}{lccc}
    \hline
    Source & $\lvert C \rvert^{2}$ & $M$ & $\kappa$ 
    \\
    \hline
    coherent state ($s = \text{coh}$)& $n^2$
    & $n^2$
    & $1$ 
    \\
    two-mode squeezed vacuum ($s= \text{TMSV}$)& $n(n+1)$
    & $n(2n+1)$
    & $\dfrac{n+1}{2n+1}$ 
    \\
     independent displaced squeezed excitation ($s = \text{sq}$)& $n_{\rm disp}^2$
    & $n^2$
    & $\left(\dfrac{n_{\rm disp}}{n}\right)^2$ 
    \\
    \hline
    \end{tabular}
\end{table*}

For TMSV excitation, \eqs~\ref{eq:correlation-law} and~\ref{eq:RN-law} yield 
\begin{align}
    R_{\rm vec}^{(\rm TMSV)}
    & =
    1 + \frac{1}{n},
    \label{eq:tmsv-Rvec}
    \\
    Y_{\rm N}^{(\rm TMSV)}
    & =
    2 + \frac{1}{n},
    \label{eq:tmsv-Y}
    \\
    R_{\rm N}^{(\rm TMSV)}(\theta)
    & =
    1 + \frac{1}{n} + \cos^{2} \chi(\theta).
    \label{eq:tmsv-RN}
\end{align}
These predictions show that enhanced two-photon correlation increases the eigenvector contribution, while the normally ordered information also depends on the molecular encoding angle $\chi$. The enhancement of the total QFI is further constrained by the common population contribution through \eq~\ref{eq:RQ-bound}.

For independent displaced squeezed excitation, the normally ordered population moment remains unchanged at fixed $n$, whereas the selected correlation decreases with the squeezed fraction. Consequently,
\begin{align}
    R_{\rm vec}^{(\rm sq)}
    & =
    \left( 1 - \frac{n_{\rm sq}}{n} \right)^{2}, 
    \label{eq:sq-Rvec}
    \\
    R_{\rm N}^{(\rm sq)}(\theta)
    & =
    \cos^{2} \chi + \left(1-\frac{n_{\rm sq}}{n}\right)^{2} \sin^{2} \chi (\theta)
    \le 1.
    \label{eq:sq-RN}
\end{align}
Thus, independent squeezing suppresses the phase-encoded contribution to $F_{\rm N}$, while leaving its amplitude-encoded component unchanged. Under the conditions of \eq~\ref{eq:RQ-bound}, the total QFI likewise cannot exceed that of the matched coherent reference at leading order.

\section{Results}
\label{sec:results}
We examine the information laws using a Frenkel exciton dimer coupled to an effective Drude-Lorentz environment. Exciton population transfer and decay during the waiting time $\tau_2$ are described by a secular population model. We retain the dominant excitation Schmidt mode pair and a single emitted signal mode, with field~1 treated as a prescribed classical drive. All source comparisons are performed in the rephasing configuration with matched temporal modes and mean photon occupations. The molecular model and numerical implementation are detailed in Sections~\ref{app:model} and~\ref{app:quantum-map} of the Supporting Information.

\subsection{Two-photon correlations and eigenvector QFI}
\label{sec:results-correlation}
\begin{figure}
    \centering
    \includegraphics{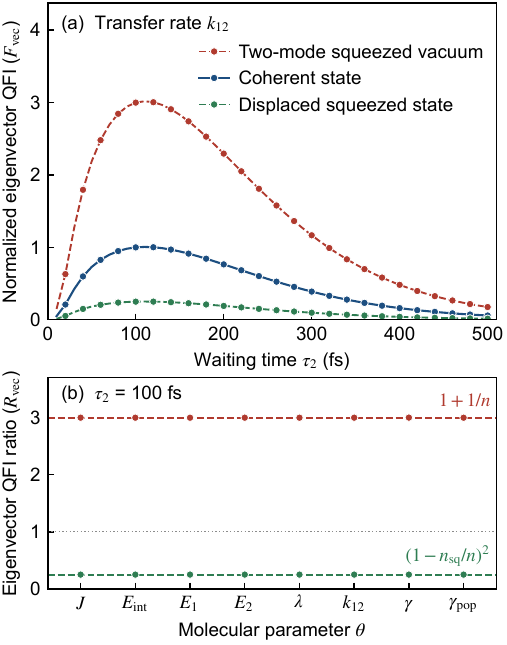}
    \caption{
    Interaction-selected correlations and eigenvector quantum Fisher information for matched excitation
    sources at $\bar{n} = 1$.
    (a) Waiting-time dependence of $F_{\rm vec}$ for the downhill population-transfer rate $k_{12}$ under coherent, two-mode squeezed vacuum (TMSV), and independent displaced squeezed excitation. All curves are normalized by the maximum coherent state value.
    (b) Eigenvector QFI ratio $R_{\rm vec}$ for eight molecular parameters at $\tau_2 = 100~\mathrm{fs}$. Symbols represent numerical results, lines indicate the leading-order prediction of \eqs~\ref{eq:tmsv-Rvec} and~\ref{eq:sq-Rvec}. The gray dotted line denotes the coherent reference, $R_{\rm vec} = 1$.
    }
    \label{fig:correlation}
\end{figure}

We first examine the dependence of the emitted signal information on the interaction-selected two-photon correlation. Figure~\ref{fig:correlation}a shows the waiting time dependence of $F_{\rm vec}$ for estimation of the downhill population transfer rate $k_{12}$ under three matched excitation sources. All calculations use a total mean photon number $\bar{n} = 1$, corresponding to $n = 0.5$ per input mode.

The three sources exhibit the same temporal profile, with a maximum at waiting time near $\tau_2 = 100\,{\rm fs}$, approximately one half of the characteristic transfer time $k_{12}^{-1}$. This maximum is consistent with the transient population distribution being most sensitive to changes in the transfer rate on this timescale. The total QFI shows the same timescale dependence discussed in Section~\ref{app:temporal-modes} of the Supporting Information, in line with recent quantum metrological analyses of pump-probe spectroscopy, where information on a relaxation rate likewise reaches its maximum at a finite delay set by the underlying relaxation dynamics~\cite{schlawin2026pp}.

Despite their identical temporal profiles, the three sources exhibit substantially different magnitudes of $F_{\rm vec}$. TMSV excitation enhances $F_{\rm vec}$ by a factor of 3 relative to the coherent reference, whereas independent displaced squeezed excitation with $n_{\rm sq} / n = 0.5$ reduces it to one quarter of the coherent value. These results agree with the correlation law in \eq~\ref{eq:correlation-law}, demonstrating that the molecular response determines the waiting time dependence of $F_{\rm vec}$, and the interaction-selected excitation correlations control its source-dependent scaling.

We further test the parameter independence of this scaling in Figure~\ref{fig:correlation}b. The eigenvector QFI ratios are evaluated for eight parameters, 
\begin{equation}
    \theta \in \{J, E_{\rm int}, E_1, E_2, \lambda, k_{12}, \gamma, \gamma_{\rm pop} \},
    \end{equation}
which characterize the electronic Hamiltonian, environmental response and population dynamics at $\tau_2 = 100~\mathrm{fs}$. The numerical results reproduce the predicted ratios $R_{\rm vec}^{(\rm TMSV)} = 3$ and $R_{\rm vec}^{(\rm sq)} = 0.25$ across the entire parameter set. This parameter-independent enhancement follows directly from the factorization of the vacuum--one-photon coherence: the molecular factor sets the absolute sensitivity but cancels in the ratio between matched excitation sources,
\begin{equation}
    S_{\rm R}^{(s)}(\omega_1, \tau_2, \omega_3; \theta)
    =
    \frac{C^{(s)}}{C^{(\rm coh)}}\,
    S_{\rm R}^{(\rm coh)}(\omega_1, \tau_2, \omega_3; \theta).
    \label{eq:shape}
\end{equation}
This indicates that the excitation source modifies only the overall complex prefactor for nonzero selected moments, leaving the normalized complex spectral shape unchanged.

This equivalence does not imply that the emitted signal states have identical QFI. While the mean-field signal is determined by the vacuum--one-photon coherence, the total QFI also depends on the eigenvalue structure through the population moment $M^{(s)}$ and additional contribution $q(\theta)$. As a consequence, the matched excitation sources can produce identical normalized complex 2D electronic spectra while encoding different amounts of molecular parameter information in the emitted quantum states. Related forms of spectroscopic equivalence between entangled photon and suitably tailored coherent probes have been identified in quantum light spectroscopy~\cite{ko2023}. A similar separation between mean signal and information appears in shift-current photodetection, where the mean photocurrent depends only on the mean photon number, whereas the QFI of the incident light is carried by the current noise~\cite{barts2026qu}.

\subsection{Normally ordered information law and total QFI bound}
\label{sec:results-law}
\begin{figure*}
    \centering
    \includegraphics{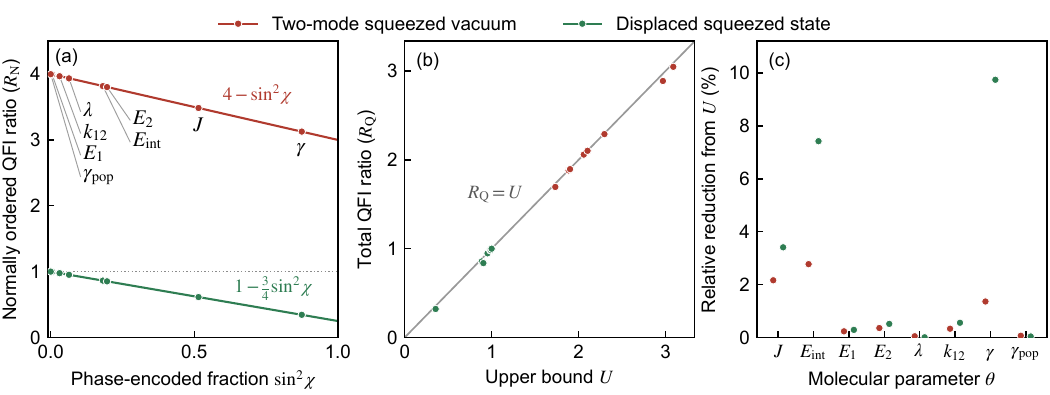}
    \caption{Normally ordered information law and total quantum Fisher information (QFI) bound for matched excitation sources at $\bar{n} = 1$ and $\tau_2=100~\mathrm{fs}$. Red and green symbols represent two-mode squeezed vacuum (TMSV) and independent displaced squeezed excitation, respectively. The latter has $n_{\rm sq} / n = 0.5$.
    (a) Normally ordered QFI ratio $R_{\rm N}$ as a function of the phase-encoded fraction $\sin^2 \chi$ for eight molecular parameters. Solid lines indicate \eq~\ref{eq:RN-law}, and symbols represent numerical results.
    (b) Total QFI ratio $R_{\rm Q}$ versus the upper bound $U$ defined in \eq~\ref{eq:U-definition}. The gray diagonal denotes $R_{\rm Q}=U$.
    (c) Relative reduction from the upper bound, $100(1-R_{\rm Q}/U)\%$, for the same molecular parameters.
    }
  \label{fig:information-law}
\end{figure*}

We next examine how the interaction-selected correlations and molecular parameter encoding jointly determine the normally ordered information, and how additional population contributions constrain the total QFI. Figure~\ref{fig:information-law} summarizes the results for eight molecular parameters at $\tau_2 = 100~\mathrm{fs}$. We consider TMSV and independent displaced squeezed excitation at $\bar{n} = 1$, with $n_{\rm sq} / n = 0.5$ for the latter. The source-dependent quantities $\kappa^{(s)}$ and $Y_{\rm N}^{(s)}$ follow from the field moments in Section~\ref{sec:source-moments}, while the encoding angle $\chi(\theta)$ is determined by the molecular response $\mathcal{G}(\theta)$. The resulting predictions contain no adjustable parameters.

Figure~\ref{fig:information-law}a shows the normally ordered QFI ratio $R_{\rm N}$ as a function of the phase-encoded fraction $\sin^2\chi(\theta)$. The numerical results follow the information law in \eq~\ref{eq:RN-law} across the considered molecular parameters, with deviations below $0.002$. For both excitation sources, $R_{\rm N}$ decreases linearly with the phase-encoded fraction, with the magnitude and slope determined by the interaction-selected field moments. This agreement confirms the combined roles of excitation correlations and molecular parameter encoding in determining the normally ordered information. 

The total QFI also depends on the additional population contribution $q(\theta)$. As shown by \eq~\ref{eq:FQ-decomposition}, the background information $F_{\rm bg}(\theta)$ is common to the matched excitation sources, while the nonnegative mixing correction $\Delta_{\rm mix}^{(s)} (\theta)$ reduces the total QFI. We define the leading-order upper bound as
\begin{equation}
    U^{(s)}(\theta)
    =
    \frac{
        R_{\rm N}^{(s)}(\theta)
        +\xi_{\rm bg}(\theta)
    }{
        1+\xi_{\rm bg}(\theta)
    },
    \label{eq:U-definition}
\end{equation}
where $\xi_{\rm bg}(\theta) = F_{\rm bg}(\theta) / F_{\rm N}^{(\rm coh)} (\theta)$. Figure~\ref{fig:information-law}b compares the calculated total QFI ratios $R_{\rm Q}^{(s)} (\theta)$ with $U^{(s)} (\theta)$. All 16 data points lie on or below the diagonal $R_{\rm Q}=U$, consistent with the leading-order bound for matched excitation sources with a common, nonnegative additional population contribution.

The deviation from the upper bound follows directly from the QFI decomposition:
\begin{equation}
    1 - \frac{R_{\rm Q}^{(s)}(\theta)}
             {U^{(s)}(\theta)}
    =
    \frac{\Delta_{\rm mix}^{(s)} (\theta)}
    {F_{\rm N}^{(s)}(\theta) + F_{\rm bg} (\theta)}.
\end{equation}
Figure~\ref{fig:information-law}c shows this relative reduction for different molecular parameters. For independent displaced squeezed excitation, the reduction reaches approximately $9.8\%$ for the bath relaxation rate $\gamma$ and $7.4\%$ for the biexciton shift $E_{\rm int}$, while remaining substantially smaller for most other parameters. The corresponding reductions under TMSV excitation are generally less pronounced.

The molecular dependence of the mixing correction can be understood from the pathway composition of the additional population contribution $q(\theta)$. With $q(\theta) = n \lvert \mathcal{G}(\theta) \rvert^{2} \beta(\theta)$ and $D^{(s)} (\theta) \propto \lvert \mathcal{G} (\theta) \rvert^{2}$, the mixing correction takes the form given in \eq~\ref{eq:app-beta-mixing}, where $\beta(\theta)$ is defined in \eq~\ref{eq:app-beta}. It is therefore controlled by the parameter dependence of pathway composition $\beta(\theta)$ and vanishes for parameters that rescale all pathways equally. The biexciton shift $E_{\rm int}$ directly modifies the ESA pathway through the energy of the doubly excited state, thereby changing its relative contribution to the total molecular response. This pathway dependence accounts for the substantial mixing-induced reduction observed for $E_{\rm int}$. More generally, the mixing correction reflects differences in the relative parameter sensitivities of the normally ordered and additional population contributions.

These results distinguish the correlation-based information law from the QFI of the complete emitted signal state. For TMSV excitation, $R_{\rm N}^{(\rm TMSV)} > 1$ provides an upper bound on the total QFI enhancement, whereas independent displaced squeezing satisfies $R_{\rm N}^{(\rm sq)} \le 1$ and cannot exceed the coherent reference QFI under the conditions of \eq~\ref{eq:RQ-bound}. The extent to which either source approaches its bound is further controlled by population mixing.

\subsection{Information redistribution under independent squeezing}
\label{sec:redistribution}
Independent displaced squeezed excitation provides a continuous means of controlling the interaction-selected coherence $\kappa$ at fixed photon occupation. As the squeezed fraction $n_{\rm sq}/n$ increases, the population moment $M^{(\rm sq)} = n^{2}$ remains unchanged, while $\kappa = (1 - n_{\rm sq}/n)^{2}$ decreases continuously. The additional population contribution $q$ is also unchanged under the matched excitation conditions. As a consequence, the leading signal population remains fixed and the vacuum--one-photon coherence is progressively suppressed. This allows us to follow how squeezing redistributes the molecular information between the eigenvector and eigenvalue contributions.

\begin{figure*}
    \centering
    \includegraphics{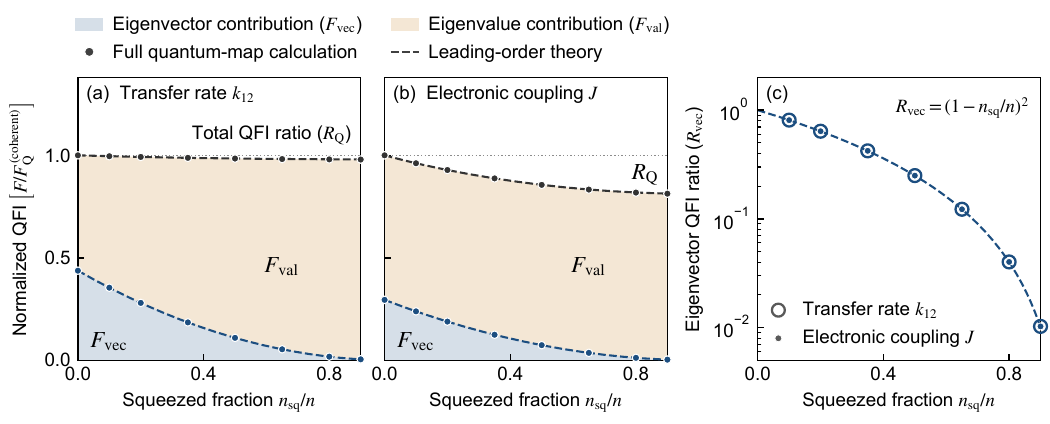}
    \caption{Information redistribution under independent displaced squeezing at $\bar{n} = 1$ and $\tau_2 = 100\,\mathrm{fs}$. 
    (a, b) Eigenvector and eigenvalue quantum Fisher information (QFI) contributions normalized by the total coherent reference QFI for  (a) the population transfer rate $k_{12}$ and (b) the electronic coupling $J$. Shaded regions indicate the two contributions; their sum gives the total QFI ratio $R_{\rm Q}$. Symbols denote the numerical results and dashed curves the leading-order predictions. The horizontal dotted line indicates the coherent reference, $R_{\rm Q} = 1$.
    (c) Eigenvector ratio $R_{\rm vec}$ for $k_{12}$ and $J$. The solid curve represents the source-only prediction of \eq~\ref{eq:sq-Rvec}.}
  \label{fig:redistribution}
\end{figure*}

Figure~\ref{fig:redistribution}a,b shows the QFI decomposition at $\tau_2 = 100~\mathrm{fs}$ for two parameters with contrasting encoding characteristics: the predominantly amplitude-encoded transfer rate $k_{12}$ and the more strongly phase-encoded electronic coupling $J$. For both parameters, the normalized fraction $F_{\rm vec}$ decreases and $F_{\rm val}$ increases with increasing squeezing. However, the total QFI responds differently. For the predominantly amplitude-encoded transfer rate $k_{12}$ (Figure~\ref{fig:redistribution}a), the reduction of $F_{\rm vec}$ is accompanied by a corresponding increase in $F_{\rm val}$, leaving the total QFI nearly unchanged. For the more strongly phase-encoded electronic coupling $J$ (Figure~\ref{fig:redistribution}b), the increase in $F_{\rm val}$ is insufficient to offset the loss of $F_{\rm vec}$, resulting in a more pronounced decrease in the total QFI. This difference arises because changes in the eigenvalue distribution cannot recover the phase sensitivity carried by the vacuum--one-photon coherence. 

The finite eigenvalue contribution already present in the coherent reference originates from the additional population $q$. The associated background information $F_{\rm bg}$ remains unchanged across the squeezing scan, reducing the relative impact of the phase-encoded information lost as $\kappa$ decreases. In the limit $\kappa \to 0$, \eq~\ref{eq:U-definition} gives 
\begin{equation}
    R_{\rm Q}^{(\rm sq)} (\theta)
    \le
    \frac{\cos^{2} \chi(\theta) + \xi_{\rm bg} (\theta)} 
         {1 + \xi_{\rm bg} (\theta)}.
\end{equation}
Population mixing further reduces the total QFI below this bound. Consequently, the background-free limit $R_{\rm Q} \to \cos^{2} \chi$ does not generally apply to the complete emitted signal state.

Figure~\ref{fig:redistribution}c isolates the eigenvector QFI ratio $R_{\rm vec}$ for the same molecular parameters. Both follow the source-only prediction of \eq~\ref{eq:correlation-law}, with $R_{\rm vec} = (1-n_{\rm sq}/n)^2$, independently of their encoding angles. At $n_{\rm sq} / n = 0.9$, the eigenvector QFI is reduced by two orders of magnitude, while the total QFI undergoes a much smaller reduction. These results demonstrate that a strong suppression of eigenvector information can be accompanied by substantial redistribution into the eigenvalue contribution, with the total information ultimately constrained by molecular parameter encoding and population mixing.

\subsection{Low-brightness scaling of QFI}
\label{sec:source}
\begin{figure}
    \centering
    \includegraphics{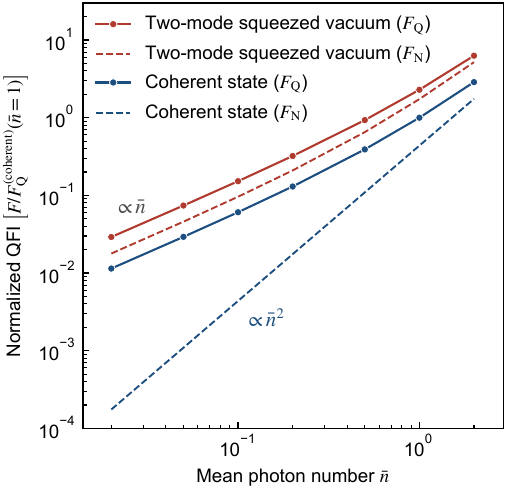}
    \caption{Brightness dependence of the normally ordered information and total quantum Fisher information (QFI) for estimation of the population transfer rate at $\tau_2 = 100~\mathrm{fs}$. Red and blue curves represent two-mode squeezed vacuum (TMSV) and coherent excitation, with matched mean photon number $\bar{n} = 2n$. Solid curves with symbols show the total QFI $F_{\rm Q}$, while dashed curves represent the normally ordered information $F_{\rm N}$. All values are normalized by the coherent reference total QFI at $\bar{n}=1$.}
    \label{fig:flux}
\end{figure}

We finally examine how the QFI of the emitted signal state depends on excitation brightness. Figure~\ref{fig:flux} compares coherent and TMSV excitation over the mean photon number range $0.02 \le \bar{n} \le 2$ for estimation of the population transfer rate $k_{12}$ at $\tau_2 = 100~\mathrm{fs}$. The normally ordered information exhibits distinct low-brightness scaling: $F_{\rm N}^{(\rm coh)} \propto \bar{n}^{2}$ for coherent excitation and $F_{\rm N}^{(\rm TMSV)} \propto \bar{n}$ for TMSV excitation. This difference follows from the interaction-selected population moments given in Section~\ref{sec:source-moments} and parallels the familiar linear-versus-quadratic photon flux scaling in entangled two-photon absorption~\cite{javanainen1990, schlawin2018, raymer2021}. This distinction highlights that the photon flux scaling of nonlinear signal generation does not directly determine the information scaling of the complete emitted optical state.

The scaling changes when the additional population contribution $q$ is included. As shown in Figure~\ref{fig:flux}, the total QFI of both coherent and TMSV excitation scales linearly with $\bar{n}$ in the low-brightness regime. In the present model, the background population $q(\theta) = n \lvert \mathcal{G}(\theta) \rvert^{2} \beta(\theta)$ scales linearly with photon number and contributes parameter-dependent eigenvalue information $F_{\rm bg}(\theta) \propto \bar{n}$. This contribution dominates the coherent total QFI at sufficiently low brightness, whereas the TMSV normally ordered information already exhibits linear scaling. Thus the distinction between linear and quadratic scaling in the normally ordered information is absent from the leading brightness dependence of the total QFI.

This change in scaling also constrains the relative information enhancement. Although the normally ordered TMSV-to-coherent ratio increases inversely with brightness, the total-QFI ratio can approach a finite, molecular parameter-dependent limit when the background carries nonzero parameter information (eq~\ref{eq:app-brightness-limit} of the Supporting Information). Moreover, the absolute QFI of both sources decreases as the excitation is weakened. The enhanced low-brightness correlations of TMSV therefore produce a larger normally ordered information contribution, while the additional population information modifies the corresponding advantage in the complete emitted signal state.

\section{Discussion}
\label{sec:discussion}
The information law clarifies which excitation-field properties are relevant to estimating a given molecular parameter. Within the present rephasing configuration, and for matched temporal modes and mean photon occupations, the amplitude-sensitive component of the normally ordered information is weighted by the joint photon-number moment $M$, whereas the phase-sensitive component is weighted by the squared magnitude of the selected pair coherence $\lvert C \rvert^2$. TMSV enhances both weights relative to coherent excitation. Independent displaced squeezing, by contrast, leaves $M$ unchanged but reduces  $\lvert C \rvert^2$, preserving the amplitude-sensitive component while suppressing the phase-sensitive component. Thus, squeezing alone does not determine the metrological benefit: the relevant correlations must be considered together with how the target parameter changes the amplitude and phase of the molecular response. Comparing the total QFI additionally requires accounting for the background population and its parameter dependence. 

Our results connect to several forms of quantum light advantage previously identified in nonlinear spectroscopy. The correlation-induced linear brightness scaling of normally ordered information parallels that of entangled two-photon absorption~\cite{javanainen1990,schlawin2018}. However, the additional population contribution $q$ modifies the corresponding information enhancement of the emitted signal state. Bounds on entangled two-photon absorption~\cite{landes2021,raymer2021} concern a different observable from the total QFI bound derived here. The distinction between the selected correlation $\lvert C \rvert^2$ and the remainder $M-\lvert C \rvert^2$ also parallels the separation of correlation-induced contributions in 2D electronic spectroscopy with entangled light~\cite{jadoun2025}. Other quantum-light advantages involve the accessible optical degrees of freedom~\cite{khan2024}, coincidence-based pathway selectivity~\cite{fujihashi2024}, spectral mapping~\cite{jadoun2026}, time- and frequency-resolved spectroscopy~\cite{alvarezMendoza2025}, or detection of nonclassical optical correlations~\cite{schlawin2026pp}. These mechanisms are distinct from the information enhancement of the retained rephasing signal state considered here.

The single Schmidt pair approximation limits the spectral equivalence established in \eq~\ref{eq:shape}. With multiple Schmidt pairs, the emitted mean field contains a weighted sum of mode-resolved molecular responses~\cite{gu2022}. Source-dependent mode weights can alter the normalized spectral shape when these responses are not mutually proportional, as illustrated by multimode quantum light spectroscopy~\cite{jadoun2026}. Extending the present analysis to multimode excitation and detection would make it possible to assess the information carried by these additional optical degrees of freedom and determine how mode-dependent molecular responses modify the source-selection criteria identified here.

The present results assume weak emission, a single retained Schmidt pair and signal mode, a prescribed classical first field, and parameter-independent excitation states. The total-QFI enhancement bound additionally requires a common, nonnegative background population.  The nonrephasing pathway instead selects $\hat{A}_2^\dagger \hat{A}_3$, for which the selected coherence vanishes for TMSV and the matched coherent reference has a nonzero normally ordered incoherent remainder. The present rephasing normalization therefore does not transfer directly to that pathway. This pathway dependence further illustrates why the relevant source correlations must be identified from the specific light–matter interaction.  

The finite-order numerical implementation does not guarantee positivity. Its small higher-order positivity violations do not change the reported QFI ratios within the precision established by the numerical convergence tests. The present third-order map gives $p_\varnothing = 0$, but does not account for possible input-independent emission contributions beyond this approximation. If such contributions are present in the detected signal mode and carry molecular parameter information, they could provide a finite common contribution to the QFI as $\bar{n} \to 0$, driving the TMSV-to-coherent ratio toward unity. The low-brightness enhancement limit discussed in Section~\ref{sec:source} therefore characterizes the present model.

\section{Conclusions}
\label{sec:conclusions}

We have established how interaction-selected two-photon correlations determine the molecular information encoded in the emitted rephasing signal state in 2D electronic spectroscopy. The selected pair coherence $C = \langle \hat{A}_2 \hat{A}_3 \rangle$ determines the vacuum--one-photon coherence, giving a parameter-independent eigenvector QFI enhancement of $1 + 1/n$ for TMSV relative to matched coherent excitation. This enhancement occurs without changing the normalized mean-field spectral shape.

The normally ordered information depends on both the selected population moment $M$ and the molecular encoding angle $\chi$. An additional population contribution, common to matched excitation sources, modifies the total-QFI enhancement, while population mixing introduces a further parameter-dependent reduction. In particular, the linear-versus-quadratic low-brightness scaling of normally ordered information does not generally persist in the total QFI of the emitted state.

These results distinguish the enhancement of nonlinear optical signals from the molecular information encoded in the corresponding quantum optical state. They also identify the selected optical correlations and accessible emission modes as central ingredients for assessing quantum-light advantages in multidimensional spectroscopy.

\section*{Supporting Information}
Additional theoretical derivations, details of the emitted-signal quantum-state construction and molecular model, and supporting numerical calculations (PDF).

\begin{acknowledgments}
A.I. is grateful to Dr.~Frank Schlawin for stimulating discussions in Okazaki, Japan. This work was supported by JSPS KAKENHI (26H02012, 26K21749 to A.I. and 23K03341 to Y.F.), and MEXT Quantum Leap Flagship Program (JPMXS0118069242 and JPMXS0120330644 to A.I.).
\end{acknowledgments}
\bibliographystyle{achemso}
\bibliography{refs}

\clearpage
\onecolumngrid

\setcounter{section}{0}
\setcounter{equation}{0}
\setcounter{figure}{0}
\setcounter{table}{0}
\renewcommand{\thesection}{S\arabic{section}}
\renewcommand{\theequation}{S\arabic{equation}}
\renewcommand{\thefigure}{S\arabic{figure}}
\renewcommand{\thetable}{S\arabic{table}}
\renewcommand{\thesubsection}{S\arabic{section}.\arabic{subsection}}

\begin{center}
\textbf{\large Supporting Information}
\end{center}

\section{Molecular model and numerical parameters}
\label{app:model}
The numerical results in the main text are obtained for the following model. We consider a Frenkel exciton dimer described by the electronic Hamiltonian
\begin{equation}
    \hat{H}_{\rm S}
    =
    \sum_{m=1}^{2} E_m \lvert m\rangle\langle m\rvert
    +
    J \left(
        \lvert 1 \rangle \langle 2 \rvert + \lvert 2 \rangle\langle 1 \rvert 
    \right)
    +
    (E_1 + E_2 + E_{\rm int}) \lvert f \rangle\langle f \rvert,
\end{equation}
where $E_m$ denotes the site excitation energy, $J$ the intermolecular electronic coupling and $E_{\rm int}$ the exciton-exciton interaction energy, which shifts the doubly excited state relative to $E_1 + E_2$. The one-exciton eigenstates are denoted by $\lvert e_1 \rangle$ and $\lvert e_2 \rangle$, with $\omega_{e_1}<\omega_{e_2}$.

Environmental broadening is represented by a common effective Drude-Lorentz spectral density with the reorganization energy $\lambda$ and the bath relaxation rate $\gamma$. Its slowly decaying high-frequency tail can introduce artifacts in the short-time bath response and complicate the Matsubara expansion at low temperatures~\cite{ishizaki2009, ishizaki2020pre}. Here it is used as a phenomenological model in the high-temperature regime.

The third-order rephasing response includes the ground-state bleach (GSB), stimulated emission (SE) and excited-state absorption (ESA) pathways, following the standard formulation of multidimensional spectroscopy~\cite{schlau2011}. During the waiting time $\tau_2$, we propagate the exciton populations using a secular Markovian rate model of incoherent transfer and excited-state decay. The downhill transfer rate $k_{12}$ describes transfer from $\lvert e_2 \rangle$ to $\lvert e_1 \rangle$, and the uphill rate $k_{21}$ is determined by detailed balance. The excited-state population decay rate $\gamma_{\rm pop}$ is assumed to be identical for both excited states, producing a common attenuation factor $\e^{-\gamma_{\rm pop} \tau_2}$ in the SE and ESA propagators.

The model parameters are listed below. For the electronic system, $E_1 = 12000~{\rm cm}^{-1}$, $E_2 = 12200~{\rm cm}^{-1}$, $J = 100~{\rm cm}^{-1}$, and $E_{\rm int} = 200~{\rm cm}^{-1}$. The environmental parameters are $\lambda = 50~{\rm cm}^{-1}$ and $\gamma^{-1} = 60~{\rm fs}$ at room temperature $300~{\rm K}$. The downhill transfer time is $k_{12}^{-1} = 200~{\rm fs}$, and the excited-state lifetime is $\gamma_{\rm pop}^{-1} = 1000~{\rm fs}$. The site transition dipoles have equal magnitudes and are mutually perpendicular, and the four dipole factors are isotropically averaged for all-parallel polarization configuration.

The coherence time is fixed at $\tau_1 = 60~\mathrm{fs}$. All three pulses have Gaussian envelopes with an intensity full width at half maximum (FWHM) of $20~\mathrm{fs}$ and carrier frequency $12100~\mathrm{cm^{-1}}$, resonant with the mean exciton energy. 

\section{Excitation fields in broadband temporal modes}
\label{app:modes}
In the main text, fields~2 and~3 are represented by a single retained Schmidt pair. Here we give the mode construction of the quantized excitation fields in orthonormal broadband temporal modes~\cite{brecht2015}, 
\begin{equation}
    \hat{A}_{jm}
    =
    \int \dd t \, \psi_{jm}^\ast (t) \hat{a}_j (t),
    \label{eq:broadband-mode}
\end{equation}
where $\hat{a}_j (t)$ denotes the time-domain annihilation operator of the $j$th excitation field. These operators satisfy $[\hat{A}_{jm}, \hat{A}_{j'm'}^\dagger] = \delta_{jj'} \delta_{mm'}$. The positive frequency field operator of the $j$th excitation field is written as
\begin{equation}
    \hat{E}_j^{(+)} (\mathbf{r}, t)
    =
    \mathcal{E}_j \, \e^{\ii \mathbf{k}_j \cdot \mathbf{r}}
    \sum_m \psi_{jm} (t) \hat{A}_{jm},
    \label{eq:field-expansion}
\end{equation}
where $\mathbf{k}_j$ is the wave vector of the $j$th pulse field. The slowly varying frequency dependent prefactor is approximated by the constant $\mathcal{E}_j$ over the pulse bandwidth.

For the correlated excitation carried by fields~2 and~3, it is convenient to choose the broadband temporal mode bases to coincide with the Schmidt modes of the two-photon amplitude~\cite{law2000, christ2011},
\begin{equation}
    \Phi(t_2, t_3) 
    =
    \sum_m \sqrt{\sigma_m} \, u_m(t_2) v_m(t_3),
\end{equation}
with $\sum_m \sigma_m = 1$, $\langle u_m \lvert u_{m'}\rangle = \delta_{mm'}$ and $\langle v_m \lvert v_{m'} \rangle = \delta_{mm'}$. We set here $\psi_{2m} (t) = u_m (t)$ and $\psi_{3m} (t) = v_m (t)$. In the single-pair approximation, only the dominant Schmidt pair ($m =1$) is retained, and we write $\hat{A}_2 = \hat{A}_{21}$ and $\hat{A}_3 = \hat{A}_{31}$. The two-photon amplitude of the correlated source is 
\begin{equation}
    \Phi (t_2, t_3) 
    \propto
    \exp \left[ - \frac{(t_2 + t_3)^2}{8 \tau_+^{2}} \right]
    \exp \left[ - \frac{(t_2 - t_3)^2}{8 \tau_-^{2}} \right]    
\end{equation}
with $\tau_+ = 20~\mathrm{fs}$ and $\tau_- = 4~\mathrm{fs}$, where $t_2$ and $t_3$ are measured from the centers of pulses~2 and~3. Its dominant Schmidt pair carries the weight $\sigma_1 = 0.56$.

Field~1 is treated as a prescribed classical drive, with its amplitude absorbed into the molecular response coefficient $\mathcal{G}(\theta)$. For the three excitation sources compared in Section~\ref{sec:results}, only the quantum state of fields~2 and~3 is varied while the retained temporal modes, matched mean photon numbers, and a common optical phase reference are held fixed. The broadband mode operators are fixed independently of the excitation state, so the source comparisons vary only the quantum states prepared in the matched optical modes.

\section{Quantum map implementation}
\label{app:quantum-map}
We employ the quantum map formulation of multidimensional spectroscopy introduced in Ref.~\cite{schlawin2025}, specialized here to the rephasing signal considered in the calculations. For fixed coherence and waiting times $\tau_1, \tau_2$, the emitted field is represented by a single temporal coordinate $t$ over the complete radiated wavepacket. When constructing the emitted quantum state, the conventional third-order response interval $\tau_3$ is not retained as an additional independent detection time variable, and is represented by this emission time coordinate $t$. The emitted signal dependence on the fixed time pair $(\tau_1, \tau_2)$ is suppressed in the following derivation.

The pathway-resolved molecular responses introduced in Section~\ref{app:model} are combined with the excitation temporal modes to construct the mode resolved emission responses $R_{m\ell}(t; \theta)$, where $m$ labels the excitation Schmidt mode and $\ell$ the Liouville pathway. The GSB, SE and ESA contributions combine coherently at the level of the emitted field amplitude, so their relative phases and signs are retained in $R_{m \ell}(t; \theta)$. To keep the joint excitation-signal Hilbert space computationally tractable, we only use the dominant Schmidt pair $m = 1$ since the Fock space dimension grows rapidly with the number of quantum modes. The emitted field is then projected onto a single retained temporal mode $\phi_0 (t)$, defining
\begin{equation}
    G_{1 \ell} (\theta)
    =
    \int \dd t \, \phi_0^\ast (t) R_{1 \ell} \, (t; \theta).
    \label{eq:Gm-appendix}
\end{equation}
The retained signal mode $\phi_0 (t)$ is obtained as the leading singular vector of the pathway- and mode-resolved emitted responses at the nominal parameter value $\theta_0$. The same mode is used at $\theta_0 + \delta\theta$ and $\theta_0 - \delta \theta$, thereby defining a fixed optical mode and phase convention for the parameter derivative and avoiding contributions associated with a parameter-dependent change of the retained basis.

In the simulation, field~1 is treated as a prescribed classical amplitude and, together with the number of emitters, is absorbed into a common interaction scale. The six rephasing diagrams of Ref~\cite{schlawin2025} comprise the ground-state bleach (GSB), stimulated emission (SE), and excited-state absorption (ESA) pathways and their Hermitian-conjugate counterparts. We use $\ell$ to label the three pathways associated with signal-photon creation, while their conjugate contributions are included explicitly in the generator below. In the ESA pathway both excitation operators act on the same side as the signal mode operator, $(\hat{L}_\ell, \hat{R}_\ell) = (\hat{X}, \hat{1})$, whereas in the SE and GSB pathways they act on opposite sides, $(\hat{L}_\ell, \hat{R}_\ell) = (\hat{A}_2, \hat{A}_3)$ and $(\hat{A}_3, \hat{A}_2)$, respectively. The resulting third-order generator acting on the joint excitation-signal space is
\begin{equation}
    \mathcal{K}_3(\theta) \hat{\rho}
    =
    -\ii \sum_\ell \Big[
     G_{1\ell} (\theta)
    \left(
        \hat{a}_{\rm s}^{\dagger} \hat{L}_\ell \hat{\rho} \hat{R}_\ell
        -
        \hat{L}_\ell \hat{\rho} \hat{R}_\ell \, \hat{a}_{\rm s}^{\dagger}
    \right)
    +
    G_{1\ell}^\ast (\theta)
    \left(
        \hat{a}_{\rm s} \hat{R}_\ell^\dagger \hat{\rho} \hat{L}_\ell^\dagger
        -
        \hat{R}_\ell^\dagger \hat{\rho} \hat{L}_\ell^\dagger \hat{a}_{\rm s}
    \right)
    \Big],
    \label{eq:K3_appendix}
\end{equation}
where $\hat{a}_{\rm s}$ annihilates a photon in the retained signal mode. For the three signal-photon-creation pathways (Figure~\ref{fig:scheme}), $\hat{R}_\ell \hat{L}_\ell = \hat{A}_2\hat{A}_3 = \hat{X}$. The total response is therefore
\begin{equation}
    \mathcal{G}(\theta)
    =
    -\ii \sum_{\ell} G_{1\ell} (\theta),
    \label{eq:pathway-sum-appendix}
\end{equation}
with
$\mathcal{G}(\theta) = \mathcal{G}_{\rm GSB}(\theta) + \mathcal{G}_{\rm SE}(\theta) + \mathcal{G}_{\rm ESA}(\theta)$. The resulting coherence is $\rho_{10}^{(s)}(\theta) = \mathcal{G}(\theta)C^{(s)}$. The one-photon population takes the closed form of \eq~\ref{eq:app-population}, with $b_2(\theta) = {\rm Re}[\mathcal{G}_{\rm SE}^{\ast}(\theta) \mathcal{G}(\theta)]$, $b_3(\theta) = {\rm Re}[\mathcal{G}_{\rm GSB}^{\ast}(\theta) \mathcal{G}(\theta)]$, and $p_\varnothing(\theta) = 0$. 

The normally ordered information $F_{\rm N}(\theta)$ is obtained from the effective signal state retaining the coherence $c = \mathcal{G}(\theta) C$ and population $P_{\rm N}(\theta) = \lvert \mathcal{G}(\theta) \rvert^{2} M$. For numerical comparison, this effective state is constructed by assigning the ESA operator placement $(\hat{L}_\ell, \hat{R}_\ell) = (\hat{X}, \hat{1})$ to all pathways while retaining their molecular coefficients. This auxiliary construction isolates the normally ordered population contribution from the full quantum map. Only the rephasing contribution is included in the numerical calculations.

For excitation source $s$, fields~2 and~3 remain quantum modes for all sources, including the coherent reference. The initial joint state is
\begin{equation}
    \hat{\rho}_{\rm in}^{(s)}
    =
    \hat{\rho}_{\rm exc}^{(s)}
    \otimes
    \lvert 0 \rangle_{\rm sig} \langle 0 \rvert,
\end{equation}
and the emitted signal state is obtained by applying the third-order map and tracing over the excitation modes,
\begin{equation}
    \label{eq:rho_sig_appendix}
    \hat{\rho}_{\rm sig}^{(s)} (\theta)
    =
    \operatorname{Tr}_{\rm exc}
    [
        \e^{\mathcal{K}_3 (\theta)}
        \hat{\rho}_{\rm in}^{(s)}
    ].
\end{equation}
The excitation and signal Hilbert spaces are represented in truncated Fock bases chosen to ensure convergence of the reported observables. The excitation modes are represented in the two-mode Fock basis $\lvert k_2, k_3 \rangle$ with $k_2 + k_3\leq B$, and the retained signal mode includes Fock states up to three photons. The budget $B$ is chosen for each figure and increased with the mean photon number, and increasing it further changes the reported ratios at the $10^{-3}$ level or below.

The QFI is evaluated from the eigendecomposition of the full retained signal mode state $\hat{\rho}_{\rm sig}^{(s)}$, and the parameter derivatives are obtained by central finite differences about $\theta_0$,
\begin{equation}
    \left.
    \frac{\partial \hat{\rho}_{\rm sig}^{(s)}}
         {\partial \theta} 
    \right|_{\theta = \theta_0}
    \approx
    \frac{
        \hat{\rho}_{\rm sig}^{(s)}(\theta_0 + \delta \theta)
        -
        \hat{\rho}_{\rm sig}^{(s)}(\theta_0 - \delta \theta)
    }
    {2 \delta \theta},
\end{equation}
using a relative step of $5 \times 10^{-3}$ and a reduced step of $2\times10^{-4}$ for the site energies $E_1$ and $E_2$. The state at the nominal parameter value is represented by the corresponding symmetric average of the two displaced states.

A common interaction rescaling,
\begin{equation}
    \label{eq:interaction-scale}
    \eta
    =
    \frac{\varepsilon}
    {\displaystyle
    \max_{\pm,\ell}
    \left \lvert
        G_{1\ell}(\theta_0 \pm \delta \theta)
    \right \rvert} ,
\end{equation}
with $\varepsilon = 0.005$ is applied to the pathway coefficients at both displaced parameter values. The same retained signal mode and interaction normalization are therefore used on both sides of the finite difference and across all matched excitation sources. This keeps the calculation within a common weak-emission regime while ensuring that differences between sources originate from their excitation state statistics. At $\tau_2 = 100~\mathrm{fs}$, the retained mode $\phi_0(t)$ carries $97.7\%$ of the squared norm of the pathway- and mode-resolved emitted responses.

Within this regime, $\hat{\rho}_{\rm sig}^{(s)}$ has small negative eigenvalues only for TMSV excitation at low brightness in Figure~\ref{fig:flux}. These eigenvalues scale as $\varepsilon^{4}$, the order neglected in the weak-emission expansion of Section~\ref{app:qfi-bound}. The smaller eigenvalue $\lambda_-$ of the vacuum--one-photon block (\eq~\ref{eq:app-small-eigenvalue}) scales as $\varepsilon^{2}$, and even the most negative eigenvalue has a magnitude more than three orders of magnitude below $\lambda_-$. Eigenvalue pairs whose sum lies at the numerical noise level, set near machine precision or by the magnitude of these negative eigenvalues, are excluded from the spectral sum for the QFI. Varying this threshold, adding a four-photon signal level, or increasing $B$ leaves $R_{\rm Q}$ unchanged.

\section{Weak-emission QFI and information bound}
\label{app:qfi-bound}
We derive the weak-emission QFI expressions and the information law introduced in Section~\ref{sec:weak-field}, and establish the bound on the total-QFI enhancement.

\subsection{Population decomposition}
For a single retained Schmidt-mode pair, the leading-order population under excitation by fields~2 and~3 is
\begin{align}
    p_{23}^{(s)} (\theta)
    & =
    P_{\rm N}^{(s)} (\theta)
    + b_2(\theta) \langle \hat{n}_2 \rangle_s
    + b_3(\theta) \langle\hat{n}_3 \rangle_s
    + p_\varnothing(\theta)
    + O(\varepsilon^4),
    \label{eq:app-population}
\end{align}
where $b_2(\theta) = \operatorname{Re} [\mathcal{G}_{\rm SE}^\ast(\theta) \mathcal{G}(\theta)]$ and $b_3(\theta) = \operatorname{Re}[\mathcal{G}_{\rm GSB}^\ast(\theta) \mathcal{G}(\theta)]$. The ESA pathway is included in the total response $\mathcal{G}$ and hence contributes to both $P_{\rm N}^{(s)}(\theta)$ and the interference coefficients $b_2(\theta)$ and $b_3(\theta)$. In the operator ordering of the collective third-order map, the single-mode occupation terms are associated with the GSB and SE amplitudes, whereas ESA enters these terms through interference with the total response. The normally ordered contribution is $P_{\rm N}^{(s)}(\theta) = \lvert \mathcal{G}(\theta) \rvert^{2} M^{(s)}$, with $M^{(s)} = \langle \hat{n}_2 \hat {n}_3\rangle_s$.

When one quantum input field is replaced by vacuum, while the marginal state of the remaining field and the interaction configuration are preserved, the populations reduce to
\begin{align}
    p_2^{(s)}(\theta)
    & =
    b_2(\theta) \langle \hat{n}_2 \rangle_s
    + p_\varnothing(\theta)
    + O(\varepsilon^4), 
    \\
    p_3^{(s)}(\theta)
    & =
    b_3(\theta) \langle \hat{n}_3 \rangle_s
    + p_\varnothing(\theta)
    + O(\varepsilon^4).
\end{align}
It follows that
\begin{equation}
    p_{23}^{(s)}(\theta) - p_2^{(s)}(\theta) -p_3^{(s)}(\theta) +p_\varnothing(\theta)
    =
    P_{\rm N}^{(s)}(\theta) + O(\varepsilon^4).
    \label{eq:app-double-difference}
\end{equation}

The remaining population contribution is therefore
\begin{equation}
    q^{(s)}(\theta)
    =
    p_{23}^{(s)}(\theta) - P_{\rm N}^{(s)}(\theta)
    =
    b_2(\theta) \langle \hat{n}_2 \rangle_s
    + b_3(\theta) \langle \hat{n}_3 \rangle_s
    + p_\varnothing(\theta)
    + O(\varepsilon^4).
    \label{eq:app-q}
\end{equation}
For excitation sources with matched mode-resolved mean photon occupations, $q^{(s)}(\theta) = q(\theta)$ to leading order.

The occupation-dependent terms arise from the pathway dependent operator placements in the quantum map. The present quantum map gives $p_\varnothing(\theta) = 0$, while additional contributions beyond this map are not determined here. The individual population terms need not be positive, but the information bound below requires the total $q(\theta)$ to be nonnegative.

\subsection{QFI of the weak-emission signal state}
We suppress the excitation-source index $s$ and consider the normalized vacuum--one-photon state of \eq~\ref{eq:signal}, with $c(\theta) = \rho_{10}(\theta)$. Its eigenvalues are
\begin{equation}
    \lambda_\pm (\theta)
    =
    \frac{1}{2} \left[
    1 \pm \sqrt{(1-2p(\theta))^{2} + 4 \lvert c(\theta) \rvert^{2}}
    \right].
    \label{eq:app-eigenvalues}
\end{equation}

In the weak-emission limit, $c(\theta) = O(\varepsilon)$ and $p(\theta) = O(\varepsilon^2)$, giving
\begin{align}
    \lambda_- (\theta)
    & =
    p(\theta) - \lvert c(\theta) \rvert^{2} + O(\varepsilon^{4}),
    \label{eq:app-small-eigenvalue}
    \\
    \lambda_+ (\theta)
    & = 
    1 - \lambda_- (\theta).
\end{align}

Specializing \eqs~\ref{eq:QFI-vec} and~\ref{eq:QFI-val} to the two eigenstates $\lvert + \rangle$ and $\lvert - \rangle$ of $\hat{\rho}$, and using $\lambda_+(\theta) + \lambda_-(\theta) = 1$, we obtain
\begin{align}
    F_{\rm vec} (\theta)
    & =
    4 \left\lvert \left\langle + \left\rvert \frac{\partial}{\partial \theta}
    \hat{\rho} \right\lvert - \right\rangle \right\rvert^{2},
    \\
    F_{\rm val} (\theta)
    & =
    \frac{\left[\dfrac{\partial}{\partial \theta} \lambda_-(\theta)\right]^{2}}
    {\lambda_-(\theta) [ 1 - \lambda_-(\theta)] }.
\end{align}
Since $\langle + \rvert \frac{\partial}{\partial \theta} \hat{\rho} \lvert - \rangle = \frac{\partial}{\partial \theta} c^\ast + O(\varepsilon^{3})$,
we obtain
\begin{align}
    F_{\rm vec} (\theta)
    & =
    4\left\lvert \frac{\partial}{\partial \theta} c(\theta) \right\rvert^{2}
    + O(\varepsilon^4),
    \label{eq:app-Fvec}
    \\
    F_{\rm val} (\theta)
    & =
    \frac{
    \left\{
        \dfrac{\partial}{\partial \theta}
        [ p(\theta) - \lvert c(\theta) \rvert^{2}]\right\}^{2}
    }{ 
        p(\theta) - \lvert c(\theta) \rvert^{2}}
    + O(\varepsilon^4).
    \label{eq:app-Fval}
\end{align}
These expansions assume a regular, positive weak-emission state. Rank-deficient cases are treated through the corresponding limits.

\subsection{Normally ordered information law}
For the normally ordered contribution, the signal coherence and population are $c^{(s)}(\theta) = \mathcal{G}(\theta) C^{(s)}$ and $P_{\rm N}^{(s)}(\theta) = \lvert \mathcal{G}(\theta) \rvert^{2} M^{(s)}$.
We introduce
\begin{equation}
    D^{(s)}(\theta)
    =
    P_{\rm N}^{(s)}(\theta) - \lvert c^{(s)}(\theta) \rvert^{2}
    =
    \lvert \mathcal{G}(\theta) \rvert^{2}
    \left( M^{(s)} - \lvert C^{(s)} \rvert^{2}\right)
    \geq 0,
    \label{eq:app-D}
\end{equation}
where the inequality follows from Cauchy--Schwarz inequality.

The normally ordered information $F_{\rm N}^{(s)}(\theta)$ is defined as the QFI of the auxiliary effective state with coherence $c^{(s)}(\theta) = \mathcal{G}(\theta) C^{(s)}$ and population $P_{\rm N}^{(s)}(\theta)$. Applying the weak-emission expansion to this state gives
\begin{align}
    F_{\rm N}^{(s)}(\theta)
    & =
    4\lvert C^{(s)} \rvert^{2}
    \left\lvert \frac{\partial}{\partial \theta} \mathcal{G}(\theta) \right\rvert^{2}
    +
    \frac{\left[\dfrac{\partial}{\partial \theta} D^{(s)(\theta)}\right]^{2}}{D^{(s)}(\theta)}
    + O(\varepsilon^4).
    \label{eq:app-FN}
\end{align}
Because the excitation moments are independent of $\theta$,
\begin{align}
    \frac{\partial}{\partial \theta} D^{(s)}(\theta)
    & =
    2 [M^{(s)} - \lvert C^{(s)} \rvert^2 ]
    \operatorname{Re}
    \left[ \mathcal{G}^\ast(\theta) \frac{\partial}{\partial \theta} \mathcal{G}(\theta) \right].
\end{align}

Using $\kappa^{(s)} = \lvert C^{(s)} \rvert^{2} / M^{(s)}$ and $\chi(\theta) = \arg\left[ \dfrac{\partial}{\partial \theta} \ln \mathcal{G}(\theta) \right]$, we find
\begin{align}
    F_{\rm N}^{(s)} (\theta)
    & =
    4M^{(s)}
    \left\lvert \frac{\partial}{\partial \theta} \mathcal{G}(\theta) \right\rvert^{2}
    [ \cos^{2} \chi(\theta) + \kappa^{(s)} \sin^{2} \chi(\theta) ]
    +O (\varepsilon^4).
    \label{eq:app-FN-law}
\end{align}
The second term in \eq~\ref{eq:app-FN} is understood by continuity when $D^{(s)}(\theta) = 0$ identically.

For the matched coherent reference, $\kappa^{(\rm coh)} = 1$. Defining $Y_{\rm N}^{(s)} = M^{(s)}/M^{(\rm coh)}$, the resulting information law is
\begin{equation}
    R_{\rm N}^{(s)} (\theta)
    =
    Y_{\rm N}^{(s)}
    [ \cos^{2} \chi(\theta) + \kappa^{(s)} \sin^{2} \chi(\theta) ].
    \label{eq:app-RN-law}
\end{equation}

\subsection{QFI decomposition and upper bound}
Including the additional population contribution $q(\theta)$, the smaller eigenvalue becomes
\begin{equation}
    \lambda_-^{(s)}(\theta)
    =
    D^{(s)}(\theta) + q(\theta) + O(\varepsilon^4).
\end{equation}
The leading eigenvalue contribution is therefore
\begin{equation}
    F_{\rm val}^{(s)}(\theta)
    =
    \frac{
    \left\{ \dfrac{\partial}{\partial \theta} [D^{(s)}(\theta) + q(\theta)] \right\}^{2}}{D^{(s)}(\theta) + q(\theta)}
    + O(\varepsilon^4).
    \label{eq:app-Fval-total}
\end{equation}

For a fixed source $s$, we write $D(\theta)=D^{(s)}(\theta)$. For positive $D(\theta)$ and $q(\theta)$, the Lagrange identity gives
\begin{align}
    \frac{\left[\dfrac{\partial}{\partial \theta} D(\theta)\right]^{2}}{D(\theta)}
    +
    \frac{\left[\dfrac{\partial}{\partial \theta} q(\theta)\right]^{2}}{q(\theta)}
    -
    \frac{\left[\dfrac{\partial}{\partial \theta}(D(\theta) + q(\theta))\right]^{2}}{D(\theta) + q(\theta)} \nonumber
    \\ =
    \frac{D(\theta) q(\theta)}{D(\theta) + q(\theta)}
    \left(\frac{\partial}{\partial \theta} \ln D(\theta)
    -
    \frac{\partial}{\partial \theta} \ln q(\theta)\right)^2
    \geq 0.
    \label{eq:app-lagrange}
\end{align}

Defining $F_{\rm bg}(\theta) = (\frac{\partial}{\partial \theta} q(\theta))^{2} / q(\theta)$ and
\begin{equation}
    \Delta_{\rm mix}^{(s)}(\theta)
    =
    \frac{D^{(s)}(\theta) q(\theta)}{D^{(s)}(\theta) + q(\theta)}
    \left[\frac{\partial}{\partial \theta} \ln \frac{D^{(s)}(\theta)}{q(\theta)} \right]^{2}
    \geq 0,
    \label{eq:app-mixing}
\end{equation}
we obtain
\begin{equation}
    F_{\rm Q}^{(s)}(\theta)
    =
    F_{\rm N}^{(s)}(\theta)
    + F_{\rm bg}(\theta)
    - \Delta_{\rm mix}^{(s)}(\theta)
    + O(\varepsilon^4).
    \label{eq:app-FQ-decomposition}
\end{equation}
Cases with $D^{(s)}(\theta) = 0$ or $q(\theta) = 0$ are treated through the corresponding limits.

For the present third-order map, $p_\varnothing(\theta) = 0$. At matched individual mode occupations $\langle \hat{n}_2 \rangle = \langle \hat{n}_3 \rangle = n$, the additional population becomes $q(\theta) = n \lvert \mathcal{G}(\theta) \rvert^{2} \beta(\theta)$, where 
\begin{equation}
    \beta(\theta)
    =
    1 - \operatorname{Re}
    \left(
        \frac{\mathcal{G}_{\rm ESA}(\theta)}{\mathcal{G}(\theta)}
    \right).
    \label{eq:app-beta}
\end{equation}
Because the excitation moments are independent of $\theta$, the common factor $\lvert \mathcal{G}(\theta) \rvert^{2}$ cancels from the logarithmic derivative of $D^{(s)}(\theta) / q(\theta)$. As a consequence,
\begin{equation}
    \Delta_{\rm mix}^{(s)}(\theta)
    =
    \frac{D^{(s)}(\theta) q(\theta)}{D^{(s)}(\theta) + q(\theta)}
    \left[
        \frac{\partial}{\partial \theta} \ln \beta(\theta)
    \right]^{2}.
    \label{eq:app-beta-mixing}
\end{equation}

For the matched coherent reference, $D^{(\rm coh)}(\theta) = 0$ identically. Consequently,
\begin{equation}
    F_{\rm Q}^{(\rm coh)}(\theta)
    =
    F_{\rm N}^{(\rm coh)}(\theta)
    + F_{\rm bg}(\theta)
    + O(\varepsilon^4).
    \label{eq:app-FQ-coh}
\end{equation}
Introducing $\xi_{\rm bg}(\theta) = F_{\rm bg}(\theta) / F_{\rm N}^{(\rm coh)}(\theta)$, the enhancement ratio becomes
\begin{align}
    R_{\rm Q}^{(s)}(\theta)
    & =
    \frac{R_{\rm N}^{(s)}(\theta) + \xi_{\rm bg} (\theta)
        - \Delta_{\rm mix}^{(s)}(\theta)/F_{\rm N}^{(\rm coh)}(\theta)
    }{
        1 + \xi_{\rm bg}(\theta)
    }
    + O(\varepsilon^2).
    \label{eq:app-RQ}
\end{align}
Since $\Delta_{\rm mix}^{(s)}(\theta) \geq 0$, the total QFI enhancement satisfies the leading-order bound
\begin{equation}
    R_{\rm Q}^{(s)}(\theta)
    \lesssim
    U^{(s)}(\theta)
    =
    \frac{R_{\rm N}^{(s)} + \xi_{\rm bg}(\theta)}{1 + \xi_{\rm bg}(\theta)},
\end{equation}
where the bound $U^{(s)}(\theta)$ lies between $R_{\rm N}^{(s)}$ and unity. For $R_{\rm N}^{(s)} \geq 1$, the nonnegative background contribution further implies $1 \leq U^{(s)}(\theta) \leq R_{\rm N}^{(s)}(\theta)$, whereas for $R_{\rm N}^{(s)}(\theta) \leq 1$, $R_{\rm N}^{(s)}(\theta) \leq U^{(s)}(\theta) \leq 1$. The bound assumes parameter-independent excitation states, a common molecular response factor $\mathcal{G}(\theta)$, and a nonnegative additional population $q(\theta)$ shared by sources with matched individual mode occupations. It applies at leading order to physical weak-emission signal states; higher-order corrections are not constrained by this inequality.

\subsection{Low brightness asymptotics}
For the present quantum map with $p_\varnothing(\theta) = 0$, the additional population satisfies $ q(\theta) = n \lvert \mathcal{G}(\theta) \rvert^{2} \beta(\theta)$. Its contribution to the total QFI is
\begin{equation}
    F_{\rm bg}(\theta)
    =
    n \frac{
        \left\{\dfrac{\partial}{\partial \theta} [\lvert \mathcal{G}(\theta) \rvert^{2} \beta(\theta) ]\right\}^{2}
    }{
        \lvert \mathcal{G}(\theta) \rvert^{2} \beta(\theta)
    }.
\end{equation}
For convenience, we define the dimensionless ratio
\begin{equation}
    w (\theta)
    =
    \frac{
        \left\{ \dfrac{\partial}{\partial \theta} [\lvert \mathcal{G}(\theta) \rvert^{2} \beta(\theta)]\right\}^{2}
    }{
        4\lvert \mathcal{G}(\theta) \rvert^{2} \beta(\theta)
        \left\lvert \dfrac{\partial}{\partial \theta} \mathcal{G}(\theta) \right\rvert^{2}
    },
\end{equation}
so that $F_{\rm bg}(\theta) = 4 n \left\lvert \dfrac{\partial}{\partial \theta} \mathcal{G}(\theta) \right\rvert^{2} w(\theta)$.
At low brightness, the normally ordered information satisfies
\begin{align}
    F_{\rm N}^{(\rm coh)}(\theta)
    & =
    4 n^{2} \left\lvert \frac{\partial}{\partial \theta} \mathcal{G}(\theta) \right\rvert^{2},
    \\
    F_{\rm N}^{(\rm TMSV)}(\theta)
    & =
    4 n \left\lvert \frac{\partial}{\partial \theta} \mathcal{G}(\theta) \right\rvert^{2} + O(n^{2}).
\end{align}
For TMSV, $D^{(\rm TMSV)}(\theta) = n^{2} \lvert \mathcal{G}(\theta) \rvert^{2}$, so the mixing correction is of order $n^{2}$. Consequently,
\begin{align}
    F_{\rm Q}^{(\rm coh)}(\theta)
    & =
    4 n \left\lvert \frac{\partial}{\partial \theta} \mathcal{G}(\theta) \right\rvert^{2} w(\theta) + O(n^{2}),
    \\
    F_{\rm Q}^{(\rm TMSV)}(\theta)
    & = 
    4 n \left\lvert \frac{\partial}{\partial \theta} \mathcal{G}(\theta) \right\rvert^{2} [1 + w(\theta)] + O(n^{2}).
\end{align}
The total QFI enhancement therefore approaches
\begin{equation}
    \lim_{n \to 0}
    R_{\rm Q}^{(\rm TMSV)}(\theta)
    =
    \frac{1 + w(\theta)}{w(\theta)},
    \qquad 
    w(\theta) > 0.
    \label{eq:app-brightness-limit}
\end{equation}
This limit assumes a nonzero, parameter sensitive background contribution and does not apply when such information is absent within the leading-order weak-emission approximation.

\section{Separating temporal-mode and photon-statistics effects}
\label{app:temporal-modes}
\begin{figure}[!htbp]
  \centering
  \includegraphics[width=0.6\linewidth]{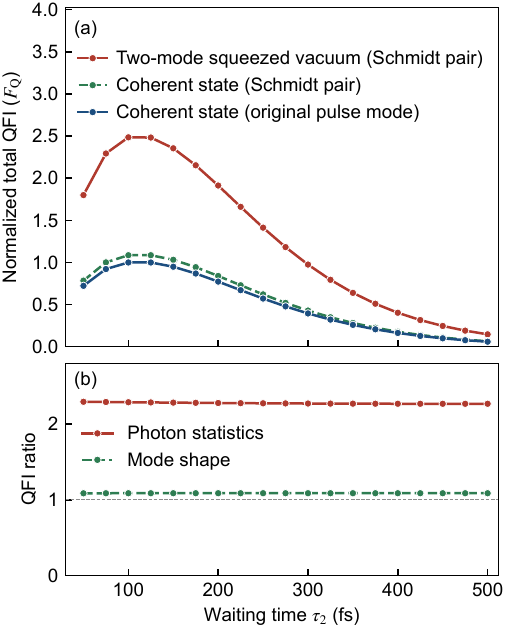}
  \caption{Separation of temporal-mode and photon-statistics effects for estimation of the downhill transfer rate $k_{12}$ at fixed total mean photon number $\bar{n} = 1$.
  (a) QFI for coherent excitation in the original pulse mode (blue), coherent excitation in the matched Schmidt mode pair (green), and TMSV in the same pair (red).
  (b) QFI ratios isolating the temporal-mode effect, $F_{\rm Q}^{(\rm coh,pair)} / F_{\rm Q}^{(\rm coh,pulse)}$ and the photon-statistics effect, $F_{\rm Q}^{(\rm TMSV,pair)} / F_{\rm Q}^{(\rm coh,pair)}$.}
  \label{fig:three}
\end{figure}

To distinguish the effects of excitation temporal modes from those of photon statistics, we compare three excitation configurations at the same total mean photon number $\bar{n} = 1$: (1) coherent excitation in the original pulse modes, (2) coherent excitation in the matched Schmidt mode pair, and (3) the TMSV excitation in the same Schmidt mode pair. Figure~\ref{fig:three} shows the total QFI for estimation of the downhill population transfer rate, with $k_{12}^{-1} = 200~\mathrm{fs}$. The waiting time scan is restricted to $\tau_2 \geq 50~{\rm fs}$, where the excitation pulses are temporally separated. Signals within or near the pulse overlap region can contain contributions from coherent artifacts, alternative pulse orderings, and perturbed free-induction decay, which complicate the interpretation of early-time population dynamics~\cite{palevcek2019, bangert2023, brosseau2023}. All three configurations exhibit nearly identical waiting time profiles, with maximal sensitivity near $\tau_2 = 100~\mathrm{fs}$.

The two ratios in Figure~\ref{fig:three}b separate the effects of mode shape and photon statistics. Changing the excitation modes from the original pulse modes to the matched Schmidt mode pair, while retaining coherent excitation, increases the QFI by only about $9 \%$. By contrast, replacing the coherent state by TMSV in the same Schmidt mode pair produces a much larger change, with $R_{\rm Q}^{(\rm TMSV)} \approx 2.3$ near the QFI maximum. Both ratios depend weakly on the waiting time $\tau_2$, despite the substantial variation of the absolute QFI. The temporal-mode effect is therefore small compared with the enhancement associated with photon statistics under the conditions considered. This comparison supports the use of matched temporal modes to isolate source-statistics effects in the main results.

\end{document}